\documentclass[longauth,traditabstract]{aa} % for the long lists of affiliations
\usepackage{graphicx}
\usepackage{txfonts}
\usepackage{natbib}

\begin{document}
   \title{The 0.5M$_J$ transiting exoplanet WASP-13b}

%   \subtitle{WASP-13b}

   \author{I. Skillen \inst{1,2}
      \and D. Pollacco\inst{2}           
      \and A. Collier Cameron\inst{3}
      \and L. Hebb\inst{3} 
      \and E. Simpson\inst{2}
      \and F. Bouchy\inst{4,5}
      \and D.J. Christian\inst{2,14}
      \and N.P. Gibson\inst{2}
      \and G. H\'ebrard\inst{4}
      \and Y.C. Joshi\inst{2}
      \and B. Loeillet\inst{12}
      \and B. Smalley\inst{9}
      \and H.C. Stempels\inst{3}
      \and R.A. Street\inst{6}
      \and S. Udry\inst{7}
      \and R.G. West\inst{8}
      \and D.R. Anderson\inst{9}
      \and S.C.C. Barros\inst{2}
      \and B. Enoch\inst{3}
      \and C.A. Haswell\inst{10}
      \and C. Hellier\inst{9}
      \and K. Horne\inst{3}
      \and J. Irwin\inst{11}
      \and F.P. Keenan\inst{2}
      \and T.A. Lister\inst{9,6}
      \and P. Maxted\inst{9}
      \and M. Mayor\inst{7}
      \and C. Moutou\inst{12}
      \and A.J. Norton\inst{10}
      \and N. Parley\inst{10,3}
      \and D. Queloz\inst{7}
      \and R. Ryans\inst{2}
      \and I. Todd\inst{2}
      \and P.J. Wheatley\inst{13}
      \and D.M. Wilson\inst{9,15} }

\institute{Isaac Newton Group of Telescopes, Apartado de Correos 321, E-38700 Santa Cruz de la       	Palma, Tenerife, Spain  \email{wji@ing.iac.es}
\and  Astrophysics Research Centre, School of Mathematics \&\ Physics, Queen's University, University Road, Belfast, BT7 1NN, UK 
\and School of Physics and Astronomy, University of St Andrews, North Haugh, St Andrews, Fife KY16 9SS, UK 
\and Institut d'Astrophysique de Paris, CNRS (UMR 7095) --  Universit\'e Pierre \&\ Marie Curie, 98$^{bis}$ bvd. Arago, 75014 Paris, France
\and Observatoire de Haute-Provence, 04870 St Michel l'Observatoire, France  
\and Las Cumbres Observatory, 6740 Cortona Dr. Suite 102, Santa Barbara, CA 93117, USA 
\and Observatoire de Gen\`eve, Universit\'e de Gen\`eve, 51 Ch. des Maillettes, 1290 Sauverny, Switzerland 
\and Department of Physics and Astronomy, University of Leicester, Leicester, LE1 7RH, UK
\and Astrophysics Group, Keele University, Staffordshire, ST5 5BG, UK 
\and Department of Physics and Astronomy, The Open University, Milton Keynes, MK7 6AA, UK 
\and Institute of Astronomy, University of Cambridge, Madingley Road, Cambridge, CB3 0HA, UK 
\and Laboratoire d'Astrophysique de Marseille, BP 8, 13376 Marseille Cedex 12, France 
\and Department of Physics, University of Warwick, Coventry CV4 7AL, UK  
\and Department of Physics and Astronomy, California State University, 18111 Nordhoff Street, Northridge, CA, USA 
\and Centre for Astrophysics and Planetary Science, School of Physical Sciences, University of Kent, Canterbury, Kent, CT2 7NH, UK }

%
%
%
%
%\and Centre for Astrophysics, Science \& Technology Research Institute, University of Hertfordshire, Hatfield, AL10 9AB, UK 

   \date{Received 04 07, 2009; accepted 05 12, 2009}

% \abstract{}{}{}{}{} 
% 5 {} token are mandatory
 
  \abstract
{We report the discovery of WASP-13b, a low-mass $ M_p = 0.46 ^{+ 0.06 }_{- 0.05 }  M_J$ transiting exoplanet with an orbital period of $4.35298 \pm 0.00004$ 
days. The transit has a depth of 9~mmag, and although our follow-up photometry 
does not allow us to constrain the impact parameter well ($0 < b < 0.46$),  
with radius in the range $R_p \sim 1.06 - 1.21 R_J$ 
 the location of WASP-13b in the mass-radius plane is nevertheless consistent with 
H/He-dominated, irradiated, low core mass and core-free theoretical models.
The G1V host star is similar to the Sun in 
mass (M$_{*} = 1.03^{+0.11}_ {- 0.09} M_{\odot}$) 
and 
metallicity ([M/H]=$0.0\pm0.2$), but is 
possibly older ($8.5^{+ 5.5 }_{- 4.9}$~Gyr).

}


   \maketitle
%
%________________________________________________________________

\section{Introduction}
The discovery of transiting planets is a prominent theme in modern 
astrophysics. Even in the era of space-borne surveys such as CoRoT \citep{barge2007}, the detection of a new transiting planet remains an 
important and celebrated discovery. In part this is because these are 
the only systems for which accurate  physical
parameters can be determined, which in turn enables their
mass-radius relationship to be used as a diagnostic to constrain
models of 
exoplanet structure and evolution. 
Currently there are some sixty transiting exoplanets known, and  
these manifest significant diversity in their physical properties. 

There are four leading ground-based transit surveys: HATNet 
\citep{bakos2004}, 
the Trans-Atlantic Exoplanet Survey \citep{dunham2004, odonovan2006}, 
WASP \citep{pollacco2006} and XO \citep{mccullough2006}. Each uses 
specialist 
instruments capable of imaging the sky over extremely large angular scales, 
and consequently are optimised to obtain high precision photometry for 
relatively bright stars. Early predictions were optimistic about the 
expected planetary yields \citep{horne2003} in such wide-angle surveys. 
As a clearer understanding 
of the effects of systematic noise on the photometry has emerged 
\citep{pont2006, smith2006}, the discovery rates %of these surveys 
now broadly reflect these predictions. For example, 
the WASP survey 
announced the discovery of thirteen transiting exoplanets over the period 
August 2007 to April 2008. While space-based surveys are superior 
to ground-based ones for the detection of small planets and long-period 
systems, ground-based surveys are likely to remain important as their 
extremely large fields-of-view make them ideal for 
surveys of bright stars, which are well suited for detailed 
follow-up observations using other facilities.

In this paper we describe the discovery of a new, relatively 
low-mass exoplanet, WASP-13b, which was detected as part of the SuperWASP-North survey.

%__________________________________________________________________

\section{Observations and Data Reduction}

\begin{figure}
\centering
\includegraphics[width=9cm, angle=-90]{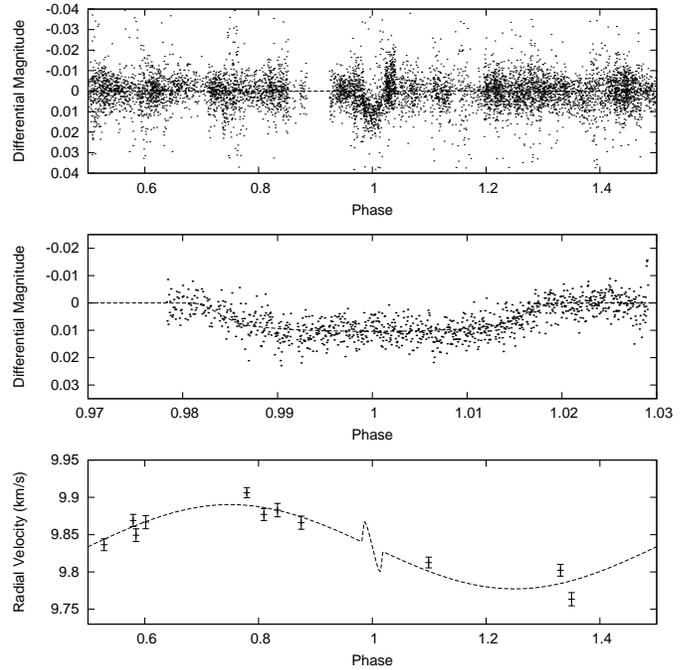}
\caption{Phased light and radial-velocity curves for WASP-13 obtained with 
SuperWASP-N (top), the JGT (middle) and SOPHIE (bottom). MCMC model 
solutions are shown as dotted lines for each data set, and the 
radial velocity  model includes the predicted 
Rossiter-McLaughlin effect assuming the upper-limit 
$v \sin i$ (4.9 km/s)}.
\label{Fig:lcs}
\end{figure}

The WASP Cameras are wide-field imaging facilities designed for the 
detection of exoplanetary transits. There are two similar facilities: 
SuperWASP-North (hereafter SuperWASP-N) on the island of La Palma in 
the Canary Islands, and WASP-South located 
at Sutherland, South Africa. The instrumentation and infrastructure used 
to obtain, store and reduce the data are described in detail in 
%\citet{pollacco2006}.  
Pollacco et al. (2006).
WASP-13 (= 1SWASP J092024.70+335256.6 = 2MASS J09202471+3352567 = USNO-B1.0 1238-0183620), a V = 10.51 G1V star in Lynx, was 
monitored with the SuperWASP-N Camera from 2006 November 27 to 2007 
April 1, during which 3329 30-second images were 
obtained with a cadence of $\sim 7$ minutes. It was identified as 
a transit candidate using the algorithm outlined by 
%\citet{cameron2007a}, 
Collier Cameron et al. (2007),
and its lightcurve is shown in Figure~\ref{Fig:lcs} (top panel). 

Further $R$-band photometry was obtained with the 0.95\,m James Gregory 
Telescope 
(JGT) located at St Andrews, Scotland, during the transit of 2008 February 16.  
The camera on this telescope consists of a 1024x1024 e2v CCD with 
an unvignetted field-of-view of $15\arcmin$ in diameter. A total of 
1047 15-second exposures was obtained in clear conditions with seeing of 
$4-5\arcsec$ throughout the night. The last 20 images in the sequence
were obtained under thick cloud cover and are not used in the final 
lightcurve.
Data were processed using the {\it Cambridge Astronomical Survey Unit}
data reduction and photometry pipeline \citep{irwinlewis}.  
Aperture photometry with an $8\arcsec$ radius aperture was performed 
on all stars in the field, and seven bright reference stars were 
selected for the differential photometry.  The flux of the combined reference 
star is dominated by a single bright star, HD80408, situated $7.6\arcmin$ 
from the target and 
approximately two magnitudes brighter.  
HD80408 (V-K=1.9) is within one spectral class of  
the planet host star (V-K=1.4).  

The transit ingress was adversely affected by 
extinction because the data were obtained at high airmass in humid conditions. 
The data display a trend with airmass which we 
removed by subtracting a  third-degree polynomial fit to the residuals 
from an initial model fit to the original differential light curve.  
There was essentially no correction to the measurements obtained 
at ${\sec Z} < 1.5$, and the amplitude of the correction 
for $1.5 < {\sec Z} < 1.8$
ranges from 
0.5 to 5~mmag. The adopted JGT 
lightcurve is shown in Figure~\ref{Fig:lcs} (middle panel).

WASP-13 was observed with the Observatoire de Haute-Provence 1.93\,m 
telescope and the SOPHIE stabilised spectrograph \citep{bouchy2006} 
during the period 2008 February 11--15, and eleven spectra were 
acquired with exposure time of 600-seconds and signal-to-noise 
$\sim 40-60$ measured in echelle order 26 at $\lambda \sim 550$ nm (SN26).  
We configured the instrument in its high efficiency mode with a 
resolution of R=40000, acquiring simultaneous star and sky spectra 
through separate fibres. Thorium-Argon calibration images were taken at 
the start and end of each night, and at 2- to 3-hourly intervals 
throughout each night.  The radial-velocity drift never exceeded 2-3 m/s, 
even on a night-to-night basis. 
Spectra were reduced with the standard SOPHIE pipeline and corrections 
applied for lunar contamination of the cross-correlation function as 
needed.

A further three spectra of WASP-13 with resolution R=46000 were acquired with the 2.5\,m Nordic Optical
Telescope (NOT) using the FIES 
echelle spectrograph, with the aim
of determining the host-star parameters (see Section~\ref{sect:stellparam}).
These spectra were extracted with the bespoke data reduction package, 
FIEStool.

%The instrument was 
%used in the simultaneous ThAr calibration mode and the data were reduced 
%with the FIEStool package. 
%A technical problem rendered these data unsuited 
%for radial velocity studies, but they are nonetheless of high quality and 
%useful in the determination of the host-star parameters (see Section~\ref{sect:stellparam}).

\begin{table}
\centering
\caption[]{Log of SOPHIE observations of WASP-13. }
\label{tab:spec_obs}
\begin{tabular}{ccrccc}
\hline\\
BJD & Phase  & SN26 & RV  & $\sigma_{\tt RV}$ & V$_{\tt span}$ \\
    &     &   & (km/s)    & (km/s)  &  (km/s) \\
\hline\\
2454508.48441  &  0.875   &   45.7  &  9.8660  &   0.0088  &   0.020  \\
2454509.45978  &  0.099   &   56.1  &  9.8126  &   0.0072  &  -0.005 \\
2454510.46914  &  0.331   &   51.7  &  9.8022  &   0.0078  &   0.017 \\
2454510.55389  &  0.350   &   45.3  &  9.7634  &   0.0088  &   0.007 \\
2454511.32838  &  0.538   &   51.1  &  9.8365  &   0.0080  &   0.016 \\
2454511.55075  &  0.579   &   50.1  &  9.8690  &   0.0082  &   0.009 \\
2454511.57290  &  0.584   &   47.3  &  9.8492  &   0.0086  &   0.032 \\
2454511.64730  &  0.602   &   46.6  &  9.8667  &   0.0088  &  -0.003 \\
2454512.42051  &  0.779   &   60.9  &  9.9062  &   0.0068  &   0.028 \\
2454512.55184  &  0.809   &   51.2  &  9.8770  &   0.0080  &   0.010 \\
2454512.65565  &  0.833   &   45.0  &  9.8829  &   0.0090  &  -0.013 \\
\hline\\
\end{tabular}
\end{table}

\section{Results}
The SuperWASP-N and JGT lightcurves show the presence of a 9~mmag  
dip of duration $\sim 3.9$~hours which repeats with a period of 
$\sim 4.35$ days. 

The radial velocities derived from the SOPHIE observations are listed in 
Table~\ref{tab:spec_obs} and plotted in the lower panel of 
Figure~\ref{Fig:lcs}. WASP-13 exhibits radial-velocity variability 
in phase with that expected from reflex motion caused by a transiting 
exoplanet. We examined the line-bisector span, V$_{\tt span}$ (Table 1),
in the manner described by 
%\citet{christian2009} 
Christian et al. (2009) 
to search for
asymmetries in spectral line profiles that could result from unresolved
binarity or indeed stellar activity. Such effects would cause the bisector
spans to vary in phase with radial velocity, but no significant correlation 
is detected. We conclude that the observed photometric
and radial-velocity variability is caused by an orbiting, planet-mass
body.
 
\subsection{Stellar and Planetary Parameters}
\label{sect:stellparam} 
 
We used the NOT echelle spectra to derive the host-star 
parameters (Table~\ref{tab:host_par}). 
Following procedures developed in our analyses of similar 
systems e.g. WASP-1 \cite{stempels2007}, we find  
$T_{\rm eff}  = 5826 \pm 100$ K,  
$\log g = 4.04 \pm 0.2$ dex 
and ${[M/H]} = 0.0 \pm 0.2$ dex, which is consistent  with solar 
metallicity. The effective temperature and near-solar mass (below) suggest a G1
spectral type. The host star has a detectable Li 6708 line from which 
we derive a lithium abundance $\log $ n(Li)  = 2.06 $\pm~0.1 $ dex. 
The rotational line profile of WASP-13 is unresolved in the FIES spectra,
and we derive an upper limit to $v \sin i$ of 4.9 km/s by subtracting in 
quadrature the macroturbulence  appropriate to $T_{\rm eff}  = 5826$ K 
(4.1 km/s) from the instrumental profile (6.4 km/s).

In addition 
to the spectrum analysis we used photometry from Tycho, $V_T=10.51$ and $(B-V)_T=0.89$, 
and 2MASS, $(V_T-H)=1.33$ and $(V_T-K)=1.39$, to estimate the effective temperature 
using the Infrared Flux Method \citep{blackwell1977}.  This yields $T_{\rm eff}  = 5935 \pm 183$ K, 
in close agreement with that obtained from the spectroscopic 
analysis. The Tycho and 2MASS colours suggest a spectral type 
of F9V  \citep{cameron2006}.

\begin{table}
\centering
\caption[]{Parameters of WASP-13.}
\label{tab:host_par}
\begin{tabular}{lcc}
\hline\\
\multicolumn{3}{c}{R.A. = 09$^h20^m24^s.70$, Dec. = $+33^\circ52^\prime56^{\prime\prime}.6$ } \\  
\\
$T_{\rm eff}$ & & $5826 \pm 100$ K \\
log ${g}$  & & $4.04 \pm 0.2$ \\
{[M/H]} & & $0.0 \pm 0.2$ \\
log n(Li) & & 2.06 $\pm~0.1$ \\
$v \sin i$ & & $<4.9$ km/s \\
Spectral type & & G1V \\
V mag & & 10.42 \\ 
Distance & & $155 \pm 18$ pc \\ 
\hline\\
\end{tabular}
\end{table}

The light and radial-velocity curves for WASP-13 were modelled simultaneously  
using the 
method described by 
%\citet{pollacco2008}.   
Pollacco et al. (2008).
The initial solution from the Monte-Carlo Markov Chain (MCMC) routine converged 
with a stellar density $\rho_{*}=0.43^{+ 0.12 }_{- 0.10 }~\rho_{\odot}$.
 To determine the mass and age of WASP13 we compared its structure and
effective temperature with the solar-metallicity stellar evolutionary
models of Girardi et al. (2000).
%\cite{girardi}. 
In Figure 2 we plot the inverse cube
root of the stellar density 
$\rho_*^{-1/3} = R_{*}/M_{*}^{1/3}$ (solar units)
against effective temperature for the model mass tracks and isochrones, and
for WASP13. We adopt this parameter space
because $\rho_{*}^{-1/3}$ unlike $R_{*}$ or luminosity, is measured directly from
the light-curve and is independent of the effective
temperature determined from the spectrum (Hebb et al. 2009). We
interpolated the evolutionary tracks and isochrones in the  $\rho_{*}^{-1/3} - \rm T_{eff}$ plane and find the 
%\cite{wasp12}, 
mass of WASP13 to be $M_{*} = 1.03^{+0.11}_ {- 0.09} M_{\odot}$ 
and its age to be $8.5^{+ 5.5 }_{- 4.9 }$~Gyr.   
Uncertainties in the derived stellar density, temperature and metallicity 
are included in the overall errors on the age and mass, but systematic 
errors due to differences between various evolutionary models are not. 
The large error in metallicity of $\pm 0.2$~dex 
contributes significantly to the uncertainty in the mass and age, and a 
more accurate spectral synthesis would improve the precision of these 
parameters.  Nevertheless, we re-ran the transit fitting code a second and 
final time, adopting an initial value for the stellar mass of 
1.03~M$_{\odot}$ and assuming a 10\%  uncertainty in this parameter.  
Our results are summarised in Table~\ref{tab:params}.

\begin{table*}
\centering
\caption[]{WASP-13 system parameters and their 1$\sigma$ error limits.} 
\label{tab:params}
\begin{tabular}{lcccl}
\hline\\
Parameter & Symbol & $b=0.46$ & $b=0.0$ &  Units \\
\hline\\
\vspace{1mm}
Transit epoch (BJD) & $ T_0  $ & $ 2454491.6161^{+ 0.0007 }_{- 0.0007 } $ &  $2454491.6161 ^{+0.0006}_{-0.0006}$ & days \\
\vspace{1mm}
Orbital period & $ P  $ & $ 4.35298^{+ 0.00004 }_{- 0.00004 } $ & $4.35298 ^{+0.00004} _{-0.00003}$ & days \\
\vspace{1mm}
Planet/star area ratio  & $ (R_p/R_s)^2 $ & $ 0.0087 ^{+ 0.0004 }_{- 0.0004 } $ & $0.0082 ^{+0.0002}_{-0.0002}$ & \\
\vspace{1mm}
Transit duration & $ t_T $ & $ 0.163 ^{+ 0.003 }_{- 0.003 } $ &  $0.160^{+0.0015}_{-0.0015} $ & days \\
\vspace{1mm}
Impact parameter & $ b $ & $ 0.46^{+ 0.13 }_{- 0.21 } $ & 0 (adopted) & $R_*$ \\
Eccentricity & $ e $ & 0 (adopted) & 0 (adopted) & \\
\vspace{1mm}
Stellar reflex velocity & $ K_1 $ & $ 0.0557^{+ 0.0054 }_{- 0.0055 } $ &  $0.0556 ^{+0.0055}_{ -0.0054}$  &    km s$^{-1}$ \\
\vspace{1mm}
Centre-of-mass velocity  & $ \gamma $ & $ 9.8340^{+ 0.0015 }_{- 0.0014 } $ &  $ 9.8340 ^{+0.0014}_{- 0.0015} $ &   km s$^{-1}$ \\
\vspace{1mm}
Orbital semimajor axis & $ a $ & $ 0.0527^{+ 0.0017 }_{- 0.0019 } $ & $0.0527 ^{+0.0018} _{-0.0020}$  &  AU \\
\vspace{1mm}
Orbital inclination & $ i $ & $ 86.9 ^{+ 1.6 }_{- 1.2 } $ & 90.0 &  degrees \\
\vspace{1mm}
Stellar mass & $ M_* $ & $ 1.03 ^{+ 0.11 }_{- 0.09 } $ &   $1.03^{+0.11}_{-0.11}$  &  $M_\odot$ \\
\vspace{1mm}
Stellar radius & $ R_* $ & $ 1.34 ^{+ 0.13 }_{- 0.11 } $ &   $1.20^{+0.04}_{-0.05}$  &  $R_\odot$ \\
\vspace{1mm}
Stellar surface gravity & $ \log g_* $ & $ 4.19 ^{+ 0.07 }_{- 0.07 } $ &    $4.29^{+0.02}_{-0.02} $  &  [cgs] \\
\vspace{1mm}
Stellar density & $ \rho_* $ & $ 0.43 ^{+ 0.12 }_{- 0.10 } $ &  $0.60^{+0.02}_{-0.02}$  &   $\rho_\odot$ \\
\vspace{1mm}
Planet radius & $ R_p $ & $ 1.21 ^{+ 0.14 }_{- 0.12 } $ & $1.06^{+0.05}_{-0.04}$ &    $R_J$ \\
\vspace{1mm}
Planet mass & $ M_p $ & $ 0.46 ^{+ 0.06 }_{- 0.05 } $ &  $0.45^{+0.05}_{-0.05}$ &   $M_J$ \\
\vspace{1mm}
Planetary surface gravity & $ \log g_p $ & $ 2.85 ^{+ 0.10 }_{- 0.10 } $ &   $3.02 ^{+0.04}_{-0.05}$   &  [cgs] \\
\vspace{1mm}
Planet density & $ \rho_p $ & $ 0.25^{+ 0.08 }_{- 0.08 } $ &   $0.39^{+0.06}_{-0.06}$  &  $\rho_J$ \\
\vspace{1mm}
Planet temperature ($A=0$)  & $ T_{eql} $ & $ 1417 ^{+ 62 }_{- 58 } $ & $1339 ^{+6}_{-6}$   &   K \\
\vspace{1mm}
Planet Safronov number & $ \Theta$ & $0.039^{+ 0.008 }_{- 0.008 }$ & $0.043^{+ 0.007 }_{- 0.007} $ & \\
\hline\\
\end{tabular}
\end{table*}

The adopted solution yields an impact parameter 
$ b = 0.46^{+ 0.12 }_{- 0.21 } $,  orbital inclination 
$i=86.9^{+1.6}_{-1.2}$ degrees 
and transit duration $\tau = 0.163 \pm 0.003 $~day, 
leading to a radius of 
R$_{*}=1.34^{+ 0.13 }_{- 0.11 }4 R_{\odot}$ for the 
1.03~M$_{\odot}$ host star. According to the stellar models, a solar metallicity star 
of this size and mass has evolved off the zero-age main sequence and 
is in the shell hydrogen burning phase of evolution with an age of 
8.5~Gyr. The Li abundance measured in the spectral synthesis also suggests
the star is several Gyr old, but it does not provide a precise 
age determination. The abundance is similar to, or slightly less than, levels found 
in open clusters with ages of 2-8~Gyr \citep{sestito}.   

Our data suggests a large stellar radius and an old age for the 
host star.  However, the JGT photometry does not  
constrain the impact parameter strongly. This affects the derived 
stellar radius and as a consequence, the derived age and  
planetary radius. Therefore, we also present a 
solution for the $b=0$ case, which gives a lower limit to the stellar 
and planetary radii.  We note that the $\chi^2$ with respect to the $b=0$
model fit is only marginally worse than the overall best fitting model.
We encourage acquisition of higher quality photometry of this 
object to enable more accurate host-star and planet radii to 
be determined.
\begin{figure}
\centering
\includegraphics[width=9cm]{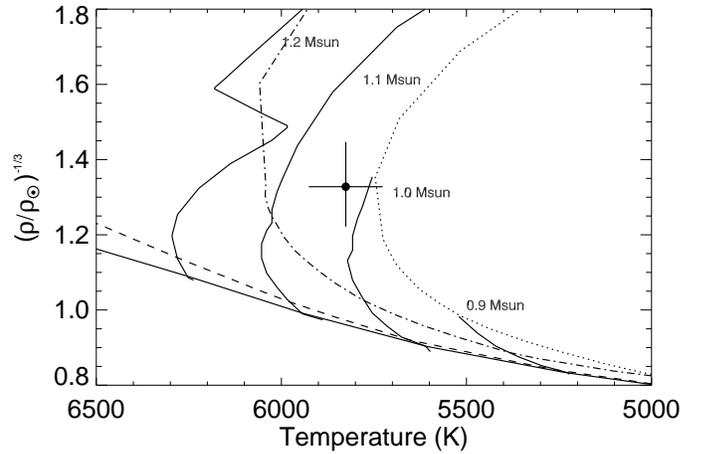}
\caption{  Location of WASP-13 in the 
$(\rho/\rho_{\odot})^{-1/3}$  
vs ${\rm T_{eff}}$ (K) plane
compared to solar-metallicity stellar evolution mass tracks from
Girardi et al. (2000). The mass tracks are labelled, and the isochrones
are 0.1 Gyr, solid; 1 Gyr, dashed; 5 Gyr, dot-dashed; 10 Gyr, dotted.
According to these models WASP-13 has a mass of 
$M_{*} = 1.03^{+0.11}_ {- 0.09} M_{\odot}$
and an age of $8.5^{+ 5.5 }_{- 4.9 } Gyr$.}
\label{fig:evol}
\end{figure}

%1.03+0.11-0.09 M_sun and
%an age of 8.5+5.5-4.9 Gyr.

%$\rho_*^{-1/3}$ versus ${\rm T_{eff}}$ for WASP-13 compared to
%solar metallicity stellar evolution tracks from \cite{girardi}.  The
%isochrones are 0.1 (solid), 1 (dashed), 5 (dot-dashed) and
%10 Gyr (dotted). The mass tracks are labelled.  According to these models,
%the star has a mass of $M_{*} = 1.03^{+0.11}_ {- 0.09} M_{\odot}$
%and an age of $8.5^{+ 5.5 }_{- 4.9 } Gyr$.}
%\label{fig:evol}
%\end{figure}
%

Although we provide only relatively weak constraints on the planet's radius, 
its mass is well constrained from the radial-velocity analysis. 
With  $ M_p = 0.46^{+ 0.05 }_{- 0.06 }  M_J$, 
WASP-13b is amongst the lowest-mass transiting exoplanets,
and with  an inflated radius in the range $R_p \sim 1.06 - 1.21 R_J$  its 
position in the mass-radius plane is broadly consistent with the
H/He-dominated, low core mass and core-free irradiated 
models of Fortney et al. (2007).

The Safronov number for WASP-13b is $\Theta = 0.039 \pm 0.008$ for
the $b=0.46$ case, and $\Theta = 0.043 \pm 0.007$ for the $b=0$ case.
This places it amongst the hotter, less-massive Class II giant
exoplanets in the classification scheme of Hansen \& Barman (2007),
who proposed that transiting exoplanets
can be distinguised into two classes according to their equilibrium
temperatures and Safronov numbers. Recently, Torres et al. (2008) reported
additional support of this dichotomy. However, Fressin et al. (2009)
suggested on the basis of simulations of a model population of stars
and exoplanets that the apparent grouping into distinct classes
is not statistically significant.
A bimodal distribution of Safronov number for giant exoplanets 
would have implications for models of their formation and evolutionary history. 
Continued discovery and parametrization of new transiting exoplanets is needed 
to confirm this hypothesis.

\begin{acknowledgements}
The WASP Consortium comprises astronomers primarily from the 
Universities of  Keele, Leicester, The Open University, Queen's University 
Belfast, the University of  St Andrews, the Isaac Newton Group (La Palma), 
the Instituto 
de  Astrof{\'i}sica de Canarias (Tenerife) and the South African 
Astronomical  Observatory. The SuperWASP-N camera is hosted  by the Isaac
Newton Group on La Palma with funding from the UK Science and 
Technology Facilities Council. We extend our thanks to the Director and 
staff of the Isaac Newton Group  
for their support of SuperWASP-N operations.
FPK would like to acknowledge 
AWE, Aldermaston for the award of a William Penney Fellowship. 

Based in part on observations made at Observatoire de Haute Provence (CNRS),
France, and on observations made with the Nordic Optical Telescope, operated
on the island of La Palma jointly by Denmark, Finland, Iceland,
Norway, and Sweden, in the Spanish Observatorio del Roque de los
Muchachos of the Instituto de Astrofisica de Canarias. 
\end{acknowledgements}


\begin{thebibliography}{}

\bibitem[Bakos et al. 2004]{bakos2004} Bakos, G., Noyes, R.W., Kov\'{a}cs, G., Stanek, K. Z., Sasselov, D. D., \& Domsa, I.
2004, PASP, 116, 266

\bibitem[Barge et al. 2007]{barge2007} Barge, P., Baglin, A., Auvergne, M. \& The CoRoT Team, 2007,
in {\it EXOPLANETS: Detection, Formation and Dynamics}, Proceedings IAU Symposium No. 249

 \bibitem[Blackwell \& Shallis 1977]{blackwell1977} Blackwell, D.E., Shallis, M.J. 1977, 
 MNRAS 180, 177 

\bibitem[Bouchy et al. 2006]{bouchy2006} Bouchy, F., The Sophie Team, 2006, in Arnold, L., Bouchy, F., Moutou, C., eds, Tenth Anniversary of 51 Peg-b: Status of and prospects for hot Jupiter studies, pp 319 -- 325.

\bibitem[Christian et al. 2009]{christian2009} Christian, D.J., et al. 2009, MNRAS, 392, 1585

\bibitem[Collier Cameron et al. 2006]{cameron2006} Collier Cameron, A., et al. 2006, 
MNRAS, 373, 799

\bibitem[Collier Cameron et al. 2007]{cameron2007a} Collier Cameron,  A., et al.  2007,
MNRAS, 380, 1230

\bibitem[Dunham et al. 2004]{dunham2004} Dunham, E.W., Mandushev, G.I., Taylor, B.W., Oetiker, B.  2004, PASP, 116, 1072

\bibitem[Fortney et al. 2007]{fortney2007} Fortney, J.J., Marley, M.S \& Barnes, J.W. 2007, ApJ, 659, 1661

\bibitem[Fressin et al. 2009]{fressin2009} Fressin, F., Guillot, T. \& Nesta, L. 2009, arXiv:0901.3083

\bibitem[Girardi et al. 2000]{girardi} Girardi, L., Bressan, A., Bertelli, G., \& Chiosi, C. 2000, A\&AS, 141, 371 

\bibitem[Hansen \& Barman 2007]{hansen2007} Hansen, B.M.S \& Barman, T. 2007, ApJ, 671, 861

\bibitem[Hebb et al. 2008]{wasp12} Hebb, L., et al.\ 2009, ApJ, 693 1920 

\bibitem[Horne 2003]{horne2003} Horne, K.D. 2003, Scientific Frontiers of Exoplanet Research, ASP Conf. 294, 361, eds. Deming \& Seager (San Francisco)

\bibitem[Irwin \& Lewis 2001]{irwinlewis} Irwin, M. \& Lewis, J. 2001, NewAR, 45, 105
   
\bibitem[McCullough et al. 2006]{mccullough2006} McCullough, P.R.  2006, 
   ApJ, 648, 1228

\bibitem[O'Donovan et al. 2006]{odonovan2006} O'Donovan, F.T., et al. 2006, ApJ, 644, 1237

      
\bibitem[Pollacco et al. 2006]{pollacco2006} Pollacco, D., et al. 2006,
   PASP, 118,  1407
   
\bibitem[Pollacco et al. 2008]{pollacco2008} Pollacco, D., et al. 2008,
   MNRAS, 385, 1576

\bibitem[Pont et al. 2006]{pont2006} Pont F., Zucker, S. \& Queloz, D. 2006, MNRAS, 373, 231

\bibitem[Sestito \& Randich 2005]{sestito} Sestito, P., \& Randich, S.\ 2005, A\&A, 442, 615

\bibitem[Smith et al. 2006]{smith2006} Smith, A.M.S., et al. 2006, MNRAS, 373, 1151
   
\bibitem[Stempels et al. 2007]{stempels2007} Stempels, H. C., Collier Cameron, A., Hebb, L., Smalley, B.,  Frandsen, S. 2007, MNRAS, 379, 773

\bibitem[Torres et al 2008]{torres2008} Torres, G., Winn, J.N., \& Holman, J. 2008, ApJ, 677, 1324


\end{thebibliography}
\end{document}